\documentclass[fleqn,12pt,twoside]{article}
\usepackage[headings]{espcrc1}

\readRCS
$Id: espcrc1.tex,v 1.2 2004/02/24 11:22:11 spepping Exp $
\usepackage{graphicx}
\usepackage{amssymb}
\usepackage[figuresright]{rotating}

\newcommand{\AmS}{{\protect\the\textfont2
  A\kern-.1667em\lower.5ex\hbox{M}\kern-.125emS}}

\title{The hadron-quark phase transition in neutron stars}

\author{G. F. Burgio\address[CT]{INFN Sezione di Catania,
        Via S. Sofia 64, I-95123 Catania, Italy}}%
       
\runtitle{The hadron-quark phase transition in neutron stars}
\runauthor{G. F. Burgio}

\begin{document}

\maketitle

\begin{abstract}
We study the hadron-quark phase transition in the interior of neutron stars
(NS). For the hadronic sector, we use a microscopic equation of state (EOS)
involving nucleons and hyperons derived within the Brueckner-Hartree-Fock 
approach. For the quark sector, we employ the MIT bag model, as well as
the Nambu--Jona-Lasinio (NJL) and the Color Dielectric (CD) models, and
find that the NS maximum masses lie in the interval between 1.5 and 1.8
solar masses.
\end{abstract}

\section{Introduction}
The appearence of quark matter in the interior of massive neutron stars
is one of the main issues in the physics of these compact objects 
\cite{pulsar}.
Calculations of NS structure, based on a microscopic nucleonic
equation of state, indicate that for the heaviest NS, close
to the maximum mass (about two solar masses), the central particle density
reaches values larger than $1/\rm fm^{3}$. In this density range, 
it can be expected that the nucleons start to loose their identity, and quark 
degrees of freedom are excited at a macroscopic level. 

The value of the maximum mass of NS is probably one of the physical 
quantities that are most sensitive to the presence of quark matter in the 
core. Unfortunately, the quark matter EOS is poorly 
known at zero temperature and at the high baryonic density 
appropriate for NS. One has, therefore, to rely on models of quark matter, 
which contain a high degree of uncertainty. 
In this paper we use a definite nucleonic EOS, which has been developed
on the basis of the Brueckner many-body theory, and three different models
for the quark EOS, respectively the MIT bag model, the Nambu--Jona-Lasinio 
and the Color Dielectric models. We compare the predictions of different 
models, and estimate the uncertainty of the results for the NS structure 
and mass.

\section{EOS of nuclear matter}

Over the last two decades the increasing interest for the equation
of state (EOS) of nuclear matter has stimulated a great deal of
theoretical activity. Phenomenological and microscopic models of
the EOS have been developed along parallel lines with complementary roles.
The latter ones include nonrelativistic
Brueckner-Hartree-Fock (BHF) theory \cite{bal} and its relativistic
counterpart, the Dirac-Brueckner (DB) theory \cite{dbhf}, the
nonrelativistic variational approach \cite{pan}, and more recently the chiral
perturbation theory \cite{chi}. In these approaches the parameters
of the interaction are fixed by the experimental nucleon-nucleon
and/or nucleon-meson scattering data. 
We have calculated the nucleonic equation of state of nuclear matter within the
BHF theory. As in all non-relativistic many-body approaches based only on
two-body forces, the EOS derived in the BHF theory fails to reproduce
some nuclear properties, such as the binding energy of light nuclei, and 
the saturation point of nuclear matter. 
The usual way of correcting this drawback is the inclusion of three-body 
forces (TBF).  In the framework of the Brueckner theory, 
we have adopted two classes of TBF, i.e. a microscopic force \cite{lom}, 
based on meson-exchange mechanisms, and the phenomenological Urbana model
\cite{uix}, 
widely used in variational calculations of finite nuclei and nuclear matter 
\cite{pan}. For details, the reader is referred to Ref.\cite{china}.
We have extended the BHF approach in a fully microscopic and 
self-consistent way, in order to describe nuclear matter containing also 
hyperons \cite{hypns}. 
We have found rather low hyperon onset densities of about 2 to 3 times 
normal nuclear matter density
for the appearance of the $\Sigma^-$ and $\Lambda$ hyperons.
(Other hyperons do not appear in the matter).

In order to study the neutron star structure, we have to calculate the 
composition and the EOS of cold, catalyzed matter, by
requiring that the neutron star contains charge neutral matter 
consisting of neutrons, protons, hyperons, and leptons ($e^-$, $\mu^-$)
in beta equilibrium. Then we compute the composition and the EOS 
in the standard way \cite{pulsar,aa}, i.e. by solving the equations
for beta-equilibrium, charge neutrality and baryon number conservation.
The inclusion of hyperons produces an EOS which is
much softer than the purely nucleonic case.
As a consequence, the maximum mass for neutron stars turns out to be
less than 1.3 solar masses \cite{hypns},
which is below the observational limit of 1.44 solar masses \cite{ht}.


\section{Quark matter}

The current theoretical description of quark matter 
is burdened with large uncertainties, and
for the time being we can only resort
to phenomenological models for EOS, 
and try to constrain them as well as possible 
by the few experimental information on high density baryonic matter.
One of these constraints is the phenomenological
observation that in heavy ion collisions at intermediate energies
($10\;{\rm MeV}/A \lesssim E/A \lesssim 200\;{\rm MeV}/A$) 
no evidence for a transition to a quark-gluon plasma has been found
up to about 3 times the saturation density $\rho_0$.
We have taken this constraint in due consideration, 
and used an extended MIT bag model \cite{chodos}
(including the possibility of a density dependent bag ``constant'') 
and the color dielectric model \cite{pi}, 
both compatible with this condition \cite{maie}. For completeness, we have also
used the Nambu--Jona-Lasinio model \cite{bub}.
\begin{figure}[t] 
\begin{center}
\includegraphics[scale=0.7,angle=270]{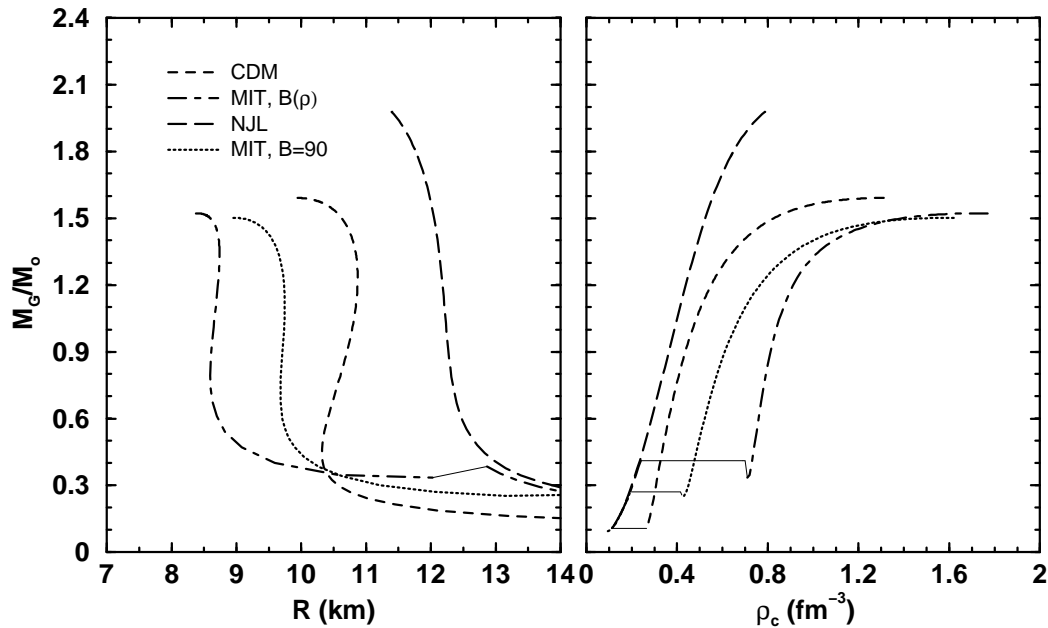}
\caption{
The gravitational mass (in units of the solar mass $M_\odot$) 
versus the radius (left panel) and the central energy density (right panel).
See text for details.}\label{f:mrq}
\end{center}
\end{figure} 
For the description of a pure quark phase inside the neutron star,
we have solved the equilibrium equations for the chemical potentials
of the different quark species, i.e. (u, d, s), along with the 
charge neutrality 
condition, and the total baryon number conservation. Hence, we have
determined the composition and the pressure of the quark phase.
In order to study the hadron-quark phase transition in neutron stars, 
one has to perform the Glendenning construction \cite{glen},
by imposing that the pressure be
the same in the two phases to ensure mechanical stability, while the 
chemical potentials of the different species are related to each other 
satisfying beta stability. This procedure
yields an EOS for the pure hadron phase, the mixed phase, and the
pure quark matter region. We have adopted a simplified method, by demanding
a sharp phase transition and performing the Maxwell construction. 
We have found that the phase transition in the extended MIT bag model 
takes place at a large baryon density, $\rho \approx 0.6\;\rm fm^{-3}$,
and at larger baryon density in the NJL model \cite{bub}. 
On the contrary, the transition density in the CD model is  
$\rho \approx 0.05\;\rm fm^{-3}$.
This implies a large difference in the structure of neutron stars.
In fact, whereas stars built with the CD model have at most a mixed phase
at low density and a pure quark core at higher density, the ones obtained
with the MIT bag model contain a hadronic phase, followed by a mixed phase
and a pure quark interior. The scenario is again different 
within the Nambu-Jona--Lasinio model, where at most a mixed phase
is present, but no pure quark phase. 

\section{Neutron star structure}

We assume that a neutron star is a spherically symmetric distribution of 
mass in hydrostatic equilibrium. 
The equilibrium configurations are obtained
by solving the Tolman-Oppenheimer-Volkoff (TOV) equations \cite{pulsar} for 
the pressure $P$ and the enclosed mass $m$,
\begin{eqnarray}
 {dp \over dr} &=& - { G m \over r^2}\, 
 { (\epsilon+p)(1+4\pi r^3 p/m) \over  1 - 2Gm/r} \:,
\\
 { dm \over dr} &=& 4\pi r^2 \epsilon \:,
\end{eqnarray}
with the newly constructed EOS for the charge neutral and beta-stable case
as input, supplemented by the EOS of the crust \cite{pulsar}.
The solutions provide information on the interior structure of a star, 
as well as the mass-radius relation, $M(R)$.
The results are shown in Fig.~\ref{f:mrq},
displaying mass-radius (left panel) and mass-central density relations (right
panel). The dashed lines represent the calculation for beta-stable quark 
matter with the CDM, whereas the dotted and dot-dashed lines denote the
results obtained with the MIT bag model (respectively for a constant bag 
constant $\rm B=90~MeV~fm^{-3}$ and a density dependent one). 
The long dashed line represents the calculations obtained within the NJL model.
We observe that the values of the maximum mass depend 
on the EOS chosen for describing quark matter, 
and lie between 1.5 and 1.97 solar masses.
We notice that the inclusion of the color superconductivity 
in the quark matter EOS built with the NJL model decreases the value of 
the maximum mass down to 1.77 $M_\odot$ \cite{bub}, thus keeping the
neutron star maximum mass well below two solar masses.
Moreover, neutron stars built with the CDM and NJL models are 
characterized by a larger radius and a smaller central density, 
whereas neutron stars constructed with the MIT bag model are more compact,
since they contain quark matter of higher density.

In conclusion, the experimental observation of a very heavy
($M \gtrsim 1.8 M_\odot$) neutron star would suggest that 
either serious problems are present for the current theoretical modelling
of the high-density phase of nuclear matter,
or that the assumptions about
the phase transition between hadron and quark phase
are substantially wrong. 
In both cases, one can expect a well defined hint on the
high density nuclear matter EOS.

\end{document}